%\input epsf
%%%%%%%%%%%%%%%%%%%%%%%%%%%%%%%%%%%%%%%%%%%%%%%%%%%%%%%%%%%%%%%%%
%                                                               %
%       FONT FAMILIES:                                          %
%                                                               %
%%%%%%%%%%%%%%%%%%%%%%%%%%%%%%%%%%%%%%%%%%%%%%%%%%%%%%%%%%%%%%%%%
%                                                               %
%       Define script letters as rsfs                           %
%               (or redefine as cal)                            %
%                                                               %
%                                                               %
%%%%%%%%%%%%%%%%%%%%%%%%%%%%%%%%%%%%%%%%%%%%%%%%%%%%%%%%%%%%%%%%%
\newfam\scrfam
\batchmode\font\tenscr=rsfs10 \errorstopmode
\ifx\tenscr\nullfont
        \message{rsfs script font not available. Replacing with calligraphic.}
        
\else   
        \font\sevenscr=rsfs7
        \font\fivescr=rsfs5
        \skewchar\tenscr='177 \skewchar\sevenscr='177 \skewchar\fivescr='177
        \textfont\scrfam=\tenscr \scriptfont\scrfam=\sevenscr
        \scriptscriptfont\scrfam=\fivescr

\fi
%%%%%%%%%%%%%%%%%%%%%%%%%%%%%%%%%%%%%%%%%%%%%%%%%%%%%%%%%%%%%%%%%
%                                                               %
%       Blackboard bold (or redefine as boldface)               %
%                                                               %
%%%%%%%%%%%%%%%%%%%%%%%%%%%%%%%%%%%%%%%%%%%%%%%%%%%%%%%%%%%%%%%%%
\newfam\msbfam
\batchmode\font\twelvemsb=msbm10 scaled\magstep1 \errorstopmode
\ifx\twelvemsb\nullfont\def\Bbb{\bf}

	\message{Blackboard bold not available. Replacing with boldface.}
\else   \catcode`\@=11
        \font\tenmsb=msbm10 \font\sevenmsb=msbm7 \font\fivemsb=msbm5
        \textfont\msbfam=\tenmsb
        \scriptfont\msbfam=\sevenmsb \scriptscriptfont\msbfam=\fivemsb
        \def\Bbb{\relax\expandafter\Bbb@}
        \def\Bbb@#1{{\Bbb@@{#1}}}
        \def\Bbb@@#1{\fam\msbfam\relax#1}
        \catcode`\@=\active

\fi
%%%%%%%%%%%%%%%%%%%%%%%%%%%%%%%%%%%%%%%%%%%%%%%%%%%%%%%%%%%%%%%%%
%                                                               %
%       MORE FONTS:                                             %
%                                                               %
%%%%%%%%%%%%%%%%%%%%%%%%%%%%%%%%%%%%%%%%%%%%%%%%%%%%%%%%%%%%%%%%%
        \font\eightrm=cmr8              \def\xrm{\eightrm}
        \font\eightbf=cmbx8             \def\xbf{\eightbf}
        \font\eightit=cmti10 at 8pt     \def\xit{\eightit}
%%%     \font\eightit=cmti8             \def\xit{\eightit}
        \font\eighttt=cmtt8             \def\xtt{\eighttt}
        \font\eightcp=cmcsc8
        \font\eighti=cmmi8              \def\xold{\eighti}
        \font\eightib=cmmib8             \def\xbold{\eightib}
        \font\teni=cmmi10               \def\old{\teni}
        \font\tencp=cmcsc10
        \font\tentt=cmtt10
        
        \font\twelvecp=cmcsc10 scaled\magstep1

	 at10pt	
	\font\twelvehelvbold=phvb at12pt
	 at14pt
	\font\sixteenhelvbold=phvb at16pt

\def\noblackbox{\overfullrule=0pt}
\noblackbox
 
%%%%%%%%%%%%%%%%%%%%%%%%%%%%%%%%%%%%%%%%%%%%%%%%%%%%%%%%%%%%%%%%%
%                                                               %
%       HEADLINE:                                               %
%                                                               %
%%%%%%%%%%%%%%%%%%%%%%%%%%%%%%%%%%%%%%%%%%%%%%%%%%%%%%%%%%%%%%%%%
\newtoks\headtext
\headline={\ifnum\pageno=1\hfill\else
{\eightcp\the\headtext}
                \dotfill{ }{\old\folio}\fi}
\def\makeheadline{\vbox to 0pt{\vss\noindent\the\headline\break
\hbox to\hsize{\hfill}}
        \vskip2\baselineskip}
%%%%%%%%%%%%%%%%%%%%%%%%%%%%%%%%%%%%%%%%%%%%%%%%%%%%%%%%%%%%%%%%%
%                                                               %
%       FOOTNOTES:                                              %
%                                                               %
%%%%%%%%%%%%%%%%%%%%%%%%%%%%%%%%%%%%%%%%%%%%%%%%%%%%%%%%%%%%%%%%%
\newcount\infootnote
\infootnote=0
\def\foot#1#2{\infootnote=1
\footnote{$\,{}^{#1}$}{\vtop{\baselineskip=.75\baselineskip
\advance\hsize by -\parindent\noindent{\xrm #2}}}\infootnote=0}
%%%%%%%%%%%%%%%%%%%%%%%%%%%%%%%%%%%%%%%%%%%%%%%%%%%%%%%%%%%%%%%%%
%                                                               %
%       REFERENCES:                                             %
%                                                               %
%%%%%%%%%%%%%%%%%%%%%%%%%%%%%%%%%%%%%%%%%%%%%%%%%%%%%%%%%%%%%%%%%
\newcount\refcount
\refcount=1
\newwrite\refwrite
\def\oldsize{\ifnum\infootnote=1\xold\else\old\fi}
\def\ref#1#2{
	\def#1{{{\oldsize\the\refcount}}\ifnum\the\refcount=1\immediate\openout\refwrite=\jobname.refs\fi\immediate\write\refwrite{\item{[{\xold\the\refcount}]} 
	#2\hfill\par\vskip-2pt}\xdef#1{{\noexpand\oldsize\the\refcount}}\global\advance\refcount by 1}
	}
\def\refout{\catcode`\@=11
        \xrm\immediate\closeout\refwrite
        \vskip2\baselineskip
        {\noindent\twelvecp References}\hfill\vskip\baselineskip
                                                %\vskip.25\baselineskip%%%%
        %\parskip=.875\parskip
        %\baselineskip=.8\baselineskip
        \baselineskip=.75\baselineskip
        \input\jobname.refs
        %\parskip=8\parskip \divide\parskip by 7
        %\baselineskip=1.25\baselineskip
        \baselineskip=4\baselineskip \divide\baselineskip by 3
        \catcode`\@=\active\rm}

\def\hepth#1{\href{http://xxx.lanl.gov/abs/hep-th/#1}{{\xtt hep-th/#1}}}
\def\JHEP#1#2#3{\href{http://jhep.sissa.it/stdsearch?paper=#1\%28#2\%29#3}{J. High Energy Phys. {\xbold #1} ({\xold#2}) {\xold#3}}}
\def\AP#1#2#3{Ann. Phys. {\xbold#1} ({\xold#2}) {\xold#3}}

\def\CQG#1#2#3{Class. Quant. Grav. {\xbold#1} ({\xold#2}) {\xold#3}}

\def\PLB#1#2#3{Phys. Lett. {\xbf B}{\xbold#1} ({\xold#2}) {\xold#3}}
\def\NPB#1#2#3{Nucl. Phys. {\xbf B}{\xbold#1} ({\xold#2}) {\xold#3}}
\def\PRD#1#2#3{Phys. Rev. {\xbf D}{\xbold#1} ({\xold#2}) {\xold#3}}

%%%%%%%%%%%%%%%%%%%%%%%%%%%%%%%%%%%%%%%%%%%%%%%%%%%%%%%%%%%%%%%%%
%                                                               %
%       SECTION NUMBERING:                                      %
%                                                               %
%%%%%%%%%%%%%%%%%%%%%%%%%%%%%%%%%%%%%%%%%%%%%%%%%%%%%%%%%%%%%%%%%
\newcount\sectioncount
\sectioncount=0
\def\section#1#2{\global\eqcount=0
	\global\subsectioncount=0
        \global\advance\sectioncount by 1
        \vskip2\baselineskip\noindent
        \line{\twelvecp\the\sectioncount. #2\hfill}
	\vskip\baselineskip\noindent
        \xdef#1{{\old\the\sectioncount}}}
\newcount\subsectioncount
\def\subsection#1#2{\global\advance\subsectioncount by 1
	\vskip.8\baselineskip\noindent
	\line{\tencp\the\sectioncount.\the\subsectioncount. #2\hfill}
	\vskip.5\baselineskip\noindent
	\xdef#1{{\old\the\sectioncount}.{\old\the\subsectioncount}}}
\newcount\appendixcount
\appendixcount=0
\def\appendix#1{\global\eqcount=0
        \global\advance\appendixcount by 1
        \vskip2\baselineskip\noindent
        \ifnum\the\appendixcount=1
        \hbox{\twelvecp Appendix A: #1\hfill}\vskip\baselineskip\noindent\fi
    \ifnum\the\appendixcount=2
        \hbox{\twelvecp Appendix B: #1\hfill}\vskip\baselineskip\noindent\fi
    \ifnum\the\appendixcount=3
        \hbox{\twelvecp Appendix C: #1\hfill}\vskip\baselineskip\noindent\fi}
\def\acknowledgements{\vskip2\baselineskip\noindent
        \underbar{\it Acknowledgements:}\ }
%%%%%%%%%%%%%%%%%%%%%%%%%%%%%%%%%%%%%%%%%%%%%%%%%%%%%%%%%%%%%%%%%
%                                                               %
%       EQUATION NUMBERING                                      %
%                                                               %
%%%%%%%%%%%%%%%%%%%%%%%%%%%%%%%%%%%%%%%%%%%%%%%%%%%%%%%%%%%%%%%%%
\newcount\eqcount
\eqcount=0
\def\Eqn#1{\global\advance\eqcount by 1
        \xdef#1{{\old\the\sectioncount}.{\old\the\eqcount}}
        \ifnum\the\appendixcount=0
                \eqno({\oldstyle\the\sectioncount}.{\oldstyle\the\eqcount})\fi
        \ifnum\the\appendixcount=1
                \eqno({\oldstyle A}.{\oldstyle\the\eqcount})\fi
        \ifnum\the\appendixcount=2
                \eqno({\oldstyle B}.{\oldstyle\the\eqcount})\fi
        \ifnum\the\appendixcount=3
                \eqno({\oldstyle C}.{\oldstyle\the\eqcount})\fi}
\def\eqn{\global\advance\eqcount by 1
        \ifnum\the\appendixcount=0
                \eqno({\oldstyle\the\sectioncount}.{\oldstyle\the\eqcount})\fi
        \ifnum\the\appendixcount=1
                \eqno({\oldstyle A}.{\oldstyle\the\eqcount})\fi
        \ifnum\the\appendixcount=2
                \eqno({\oldstyle B}.{\oldstyle\the\eqcount})\fi
        \ifnum\the\appendixcount=3
                \eqno({\oldstyle C}.{\oldstyle\the\eqcount})\fi}
\def\multi{\global\advance\eqcount by 1}
\def\multieq#1#2{\xdef#1{{\old\the\eqcount#2}}
        \eqno{({\oldstyle\the\eqcount#2})}}
%%%%%%%%%%%%%%%%%%%%%%%%%%%%%%%%%%%%%%%%%%%%%%%%%%%%%%%%%%%%%%%%%
%                                                               %
%       Hyperrefs:                                        	%
%                                                               %
%%%%%%%%%%%%%%%%%%%%%%%%%%%%%%%%%%%%%%%%%%%%%%%%%%%%%%%%%%%%%%%%%
\newtoks\url
\def\Href#1#2{\catcode`\#=12\url={#1}\catcode`\#=\active#2}
\def\href#1#2{{#2}}

%%%%%%%%%%%%%%%%%%%%%%%%%%%%%%%%%%%%%%%%%%%%%%%%%%%%%%%%%%%%%%%%%
%                                                               %
%       FORMAT:                                                 %
%                                                               %
%%%%%%%%%%%%%%%%%%%%%%%%%%%%%%%%%%%%%%%%%%%%%%%%%%%%%%%%%%%%%%%%%
\parskip=3.5pt plus .3pt minus .3pt
\baselineskip=14pt plus .1pt minus .05pt
\lineskip=.5pt plus .05pt minus .05pt
\lineskiplimit=.5pt
\abovedisplayskip=18pt plus 4pt minus 2pt
\belowdisplayskip=\abovedisplayskip
\hsize=14cm
\vsize=19cm
\hoffset=1.5cm
\voffset=1.8cm
\frenchspacing
\footline={}
%%%%%%%%%%%%%%%%%%%%%%%%%%%%%%%%%%%%%%%%%%%%%%%%%%%%%%%%%%%%%%%%%
%                                                               %
%       VARIOUS DEFINITIONS                                     %
%                                                               %
%%%%%%%%%%%%%%%%%%%%%%%%%%%%%%%%%%%%%%%%%%%%%%%%%%%%%%%%%%%%%%%%%
\def\ss{\scriptstyle}

\def\*{\partial}
\def\punkt{\,\,.}
\def\komma{\,\,,}

\def\+{\!+\!}
\def\={\!=\!}
\def\small#1{{\hbox{$#1$}}}
\def\half{\small{1\over2}}
\def\fraction#1{\small{1\over#1}}
\def\fr{\fraction}
\def\Fraction#1#2{\small{#1\over#2}}
\def\Fr{\Fraction}
\def\tr{\hbox{\rm tr}}
\def\eg{{\tenit e.g.}}
\def\ie{{\tenit i.e.}}

\def\nlni{\hfill\break}

\def\a{\alpha}
\def\b{\beta}
\def\d{\delta}

\def\c{\gamma}

\def\G{\Gamma}

\def\O{\Omega}
\def\o{\omega}
\def\tX{\tilde X}

\def\ggr{\!\cdot\!}

%%%%%%%%%%%%%%%%%%%%%%%%%%%%%%%%%%%%%%%%%%%%%%%%%%%%%%%%%%%%%%%%%%%%%%%%%
%									%
%									%
%	THE PAPER							%
%									%
%									%
%%%%%%%%%%%%%%%%%%%%%%%%%%%%%%%%%%%%%%%%%%%%%%%%%%%%%%%%%%%%%%%%%%%%%%%%%
\headtext={Cederwall, Gran, Nielsen, Nilsson: ``Manifestly Supersymmetric 
M-Theory''}

\null\vskip-2cm
%\line{
%\epsfysize=1.7cm
%\epsffile{chalmers-gu-logos.eps}
%\hfill}
%\vskip-1.7cm
\line{\hfill G\"oteborg ITP preprint}
\line{\hfill\tt hep-th/0007035}
\line{\hfill June, {\old2000}}
\line{\hrulefill}

\vfill

\centerline{\sixteenhelvbold Manifestly Supersymmetric 
M-Theory}

\vskip1.6cm

\centerline{\twelvehelvbold Martin Cederwall, Ulf Gran, Mikkel Nielsen
	and Bengt E.W. Nilsson}

\vskip.8cm

\centerline{\it Institute for Theoretical Physics}
\centerline{\it G\"oteborg University and Chalmers University of Technology }
\centerline{\it SE-412 96 G\"oteborg, Sweden}

\vskip1.6cm

%{\narrower
\noindent\underbar{Abstract:} In this paper, the low-energy 
effective dynamics of M-theory, 
eleven-dimen\-sional supergravity, is taken off-shell in a manifestly 
supersymmetric formulation. We show that a previously proposed
relaxation of the superspace 
torsion constraints does indeed accommodate a current
supermultiplet which lifts the equations of motion corresponding
to the ordinary second order derivative supergravity lagrangian. 
%Whether the
%auxiliary fields obtained this way can be used to construct an off-shell 
%lagrangian is not yet known. 
We comment on
the relation and application of this completely general formalism 
to higher-derivative ($R^4$) corrections.
Some details of the calculation are saved for a later publication.
%\smallskip}
\vfill

\line{\hrulefill}
\catcode`\@=11
\line{\tentt martin.cederwall@fy.chalmers.se\hfill}
\line{\tentt gran@fy.chalmers.se\hfill}
\line{\tentt mikkel@fy.chalmers.se\hfill}
\line{\tentt bengt.nilsson@fy.chalmers.se\hfill}
\catcode`\@=\active

\eject

\ref\VafaWitten{C.~Vafa and E.~Witten, 
{\xit ``A one loop test of string duality''}, \NPB{447}{1995}{261} 
[\hepth{9505053}].}

\ref\Duffoneloop{M.J.~Duff, J.T.~Liu and R.~Minasian, {\xit 
``Eleven-dimensional origin of string-string duality: a one loop test''}, 
\NPB{452}{1995}{261} [\hepth{9506126}].}

\ref\Greenoneloop{M.B.~Green, M.~Gutperle and P.~Vanhove, {\xit ``One-loop 
in eleven dimensions''}, \PLB{409}{1997}{177} [\hepth{9706175}].}

\ref\RussoTseytlinoneloop{J.G.~Russo and A.A.~Tseytlin, 
{\xit ``One-loop four-graviton amplitude in eleven-dimensional supergravity''},
\NPB{508}{1997}{245} [\hepth{9707134}].}

\ref\Greentwoloop{M.B.~Green, H.-h.~Kwon and P.~Vanhove, 
{\xit ``Two loops in eleven dimensions''}, 
\PRD{61}{2000}{104010} [\hepth{9910055}].}

\ref\GreenGutKwona{M.B.~Green, M. Gutperle and H.-h.~Kwon,
{\xit ``Light-cone quantum mechanics of the eleven-dimensional
superparticle''},
\JHEP{08}{1999}{012} [\hepth{9907155}].}

\ref\GreenGutKwonb{M.B.~Green, M. Gutperle and H.-h.~Kwon,
{\xit ``$\ss\lambda^{16}$ and related terms in M-theory on $\ss T^2$''},
\PLB{421}{1998}{149} [\hepth{9710151}].}

\ref\DeRooI{M.~de Roo, H.~Suelmann and A.~Wiedemann,
{\xit ``The supersymmetric effective action of the heterotic string''},
\NPB{405}{1993}{326} [\hepth{9210099}].}

\ref\BNVI{B.E.W.~Nilsson, 
\xit ``Off-shell fields for the 10-dimensional supersymmetric 
Yang--Mills theory'', \xrm G\"oteborg-ITP-{\xold81}-{\xold6}.}

\ref\BNXXIV{B.E.W.~Nilsson, 
{\xit ''Pure spinors as auxiliary fields in the ten-dimensional 
supersymmetric Yang--Mills theory''},
\CQG3{1986}{{\xrm L}41}.}

\ref\BNLXVI{B.E.W.~Nilsson, {\xit ``A supersymmetric approach to branes and 
supergravity''}, \xrm in ``Theory of elementary particles'', Proc. of the
{\xold31}st international symposium Ahrenshoop, September {\xold2}-{\xold6}, 
{\xold1997}, Buckow,
Eds H. Dorn et al. (Wiley-VCH {\xold1998}), 
G\"oteborg-ITP-{\xold98}-{\xold09} [\hepth{0007017}].}

\ref\HNvN{P.~Howe, H.~Nicolai and A.~van Proeyen,
{\xit ``Auxiliary fields and a superspace lagrangian for 
linearized ten-dimensional supergravity''},
\PLB{112}{1982}{446}.}

\ref\KalloshMarstrand{R. Kallosh, {\xit ``Strings and superspace''},
\xrm in ``Unification of fundamental interactions'', Proc. of Nobel Symposium
{\xold67}, Marstrand, Sweden, June {\xold2}-{\xold7}, {\xold1998}, 
Eds L. Brink et al, Physica Scripta {\xbf T15}. }

\ref\BNXXXIV{B.E.W.~Nilsson and A.~Tollst\'en, 
{\xit ``Supersymmetrization of $\ss\zeta(3)R^4$ in superstring theory''},
\PLB{181}{1986}{63}.}

\ref\CGNNIII{M. Cederwall, U. Gran, M. Nielsen and B.E.W. Nilsson, 
{\xit in preparation}.}

\ref\PVW{K.~Peeters, P.~Vanhove and A.~Westerberg,
{\xit in preparation}.
%{\xit ``''},
%\hepth{}.
}

\ref\SiegelRocek{M.~Ro\v cek and W.~Siegel,
{\xit ``On off-shell multiplets''}, 
\PLB{105}{1981}{275}.}

\ref\BSTMii{E. Bergshoeff, E. Sezgin and P.K. Townsend, 
{\xit ``Supermembranes and eleven-dimensional supergravity''},
\PLB{189}{1987}{75};
{\xit ``Properties of the eleven-dimensional supermembrane theory''},
\AP{185}{1988}{330}.}

\ref\DuffStelleM{M.J. Duff, P.S. Howe, T. Inami and K.S. Stelle,
{\xit ``Superstrings in D=10 from supermembranes in D=11''},
\PLB{191}{1987}{70}.}

\ref\CNS{M. Cederwall, B.E.W. Nilsson and P. Sundell,
{\xit ``An action for the super-5-brane in D=11 supergravity''},
\JHEP{04}{1998}{007}, [\hepth{9712059}].}

\ref\Dthree{M. Cederwall, A. von Gussich, B.E.W. Nilsson
 and A. Westerberg,
{\xit ``The Dirichlet super-three-brane in ten-dimensional type IIB 
supergravity''},
\NPB{490}{1997}{179} [\hepth{9610148}].}

\ref\SchwarzDbrane{M. Aganagi\'c, C. Popescu, J.H. Schwarz,
{\xit ``D-brane actions with local kappa symmetry''},
\PLB{393}{1997}{311} [\hepth{9610249}].}

\ref\Dp{M. Cederwall, A. von Gussich, B.E.W. Nilsson, P. Sundell 
and A. Westerberg,
{\xit ``The Dirichlet super-p-branes in ten-dimensional type IIA and IIB 
supergravity''},
\NPB{490}{1997}{163} [\hepth{9611159}].}

\ref\BTDbrane{E. Bergshoeff and P.K. Townsend, 
{\xit ``Super D-branes''},
\NPB{490}{1997}{145} [\hepth{9711173}].}

\ref\BBG{C.P.~Bachas, P. Bain and M.B.~Green, 
{\xit ``Curvature terms in the D-brane actions and their M-theory origin''},
\JHEP{05}{1999}{11} [\hepth{9903210}].}

\ref\GreenSethi{M.B.~Green and S.~Sethi, 
{\xit ``Supersymmetry constraints on type IIB supergravity''},
\PRD{59}{1999}{046006} [\hepth{9808061}].}

\ref\BNXXVIII{B.E.W. Nilsson and A.K. Tollst\'en,
{\xit ``Superspace formulation of the ten-dimensional coupled 
Einstein--Yang--Mills system''},
\PLB{171}{1986}{212}. }

\ref\PTafs{I. Pesando and A.K. Tollst\'en,
{\xit ``Heterotic instantons and solitons in anomaly-free supergravity''},
\PLB{274}{1992}{374}. }

\ref\HoweWeyl{P.~Howe,
{\xit ``Weyl superspace''},
\PLB{415}{1997}{149} [\hepth{9707184}].}

\ref\CL{A.~Candiello and K.~Lechner,
{\xit ``Duality in supergravity theories''},
\NPB{412}{1994}{479}.}

%\ref\Dragon{N. Dragon, 
%{\xit ``Torsion and curvature in extended supergravity''},
%\ZP{C2}{1979}{29}.} 

\ref\ElevenSG{E. Cremmer, B. Julia and J. Sherk, 
{\xit ``Supergravity theory in eleven-dimensions''},
\PLB{76}{1978}{409}.}

\ref\ElevenSSSG{L. Brink and P. Howe, 
{\xit ``Eleven-dimensional supergravity on the mass-shell in superspace''},
\PLB{91}{1980}{384};
E. Cremmer and S. Ferrara,
{\xit ``Formulation of eleven-dimensional supergravity in superspace''},
\PLB{91}{1980}{61}.}

\ref\LiE{A.M. Cohen, M. van Leeuwen and B. Lisser, 
LiE v. {\xold2}.{\xold2} ({\xold1998}), 
\nlni http://wallis.univ-poitiers.fr/\~{}maavl/LiE/} 

\ref\BNVII{B.E.W. Nilsson, 
{\xit ``Simple 10-dimensional supergravity in superspace''},
\NPB{188}{1981}{176}.}

\ref\BNXXXI{B.E.W. Nilsson, 
{\xit ``Off-shell d= 10, N=1 Poincar\'e  supergravity and the embeddibility
of higher derivative field theories in superspace''},
\PLB{175}{1986}{319}.}

\ref\ConvConstr{S.J.~Gates, K.S.~Stelle and P.C.~West,
{\xit ``Algebraic origins of superspace constraints in supergravity''},
\NPB{169}{1980}{347}; 
S.J. Gates and W. Siegel, 
{\xit ``Understanding constraints in superspace formulation of supergravity''},
\NPB{163}{1980}{519}.}

\ref\Ulf{U. Gran, {\xit in preparation}.}

\ref\GatesNishino{S.J.~Gates and H. Nishino,
{\xit ``Toward an off-shell 11-D supergravity limit of M theory''},
\PLB{388}{1996}{504} [\hepth{9602011}].}

\ref\Wyllard{N. Wyllard,
{\xit 
``Derivative corrections to D-brane actions with constant background fields''},
\hepth{0008125}.}

\ref\TaylorRivelles{V.O. Rivelles and J.G. Taylor, 
{\xit ``Off-shell no go theorems for higher dimensional supersymmetries and
supergravities''},
\PLB{121}{1983}{37}.}

\ref\HSW{P.S. Howe, E. Sezgin and P.C. West,  
{\xit ``Aspects of superembeddings''},
\hepth{9705093}.}

\section\introduction{Introduction}One approach to probing M-theory 
at short distances is to consider the
effective action beyond its lowest order approximation given by the
second order (in $\#\hbox{(derivatives)}+\half\#\hbox{(fermions)}$)
action [\ElevenSG]
$$
\eqalign{
-2\kappa^2S=&\int d^{11}x\sqrt{-g}\left(R+\fr{2\cdot4!}H^{mnpq}H_{mnpq}\right)
+\fr6\int C\wedge H\wedge H \cr
&+\hbox{terms with fermions}\komma\cr
}
\Eqn\Lagrangian
$$
and investigate the higher-derivative corrections generated by the
microscopic theory.  Such  corrections at order $R^2$ and $R^4$ have been
extensively discussed in the literature, primarily in the context of
string theory and ten-dimensional effective actions, but also in the
eleven-dimensional context relevant to M-theory. The existence of
these terms can be inferred by a variety of means in string theory,
while in M-theory one must rely on anomaly cancellation arguments 
[\VafaWitten,\Duffoneloop], or 
(superparticle) loop calculations 
[\Greenoneloop,\RussoTseytlinoneloop,\GreenGutKwonb,\GreenGutKwona,\Greentwoloop] together with the connection to string
theory and its effective action via dimensional reduction.

The methods used so far to deduce the existence of \eg\ $R^4$ terms in
eleven dimensions produce only isolated terms out of a large number of
terms making up the complete superinvariant that it belongs to. It is
of interest to have a better understanding of these superinvariants,
and there has consequently been a lot of work invested into the
supersymmetrisation of the isolated terms. In particular, $R^2$ and
$R^4$ terms in ten dimensions were considered already some time ago,
see ref. [\DeRooI] and references therein.
More recently also the $R^4$ term in eleven dimensions has been
investigated [\PVW] including a detailed study of superinvariants.

For these purposes it would be interesting to develop 
methods [\BNLXVI] based on
superspace in eleven dimensions [\ElevenSSSG] that would incorporate
supersymmetry in a manifest way. Although not yet developed into an
easily applicable formalism, $N$=1 supergravity in ten dimensions has been
constructed off-shell in terms of a linearised superspace lagrangian
[\HNvN],
including some superinvariants [\KalloshMarstrand,\BNXXXIV], 
and should in principle lend itself to a
complete analysis of superinvariants and deduction of the
corresponding higher-derivative terms in ordinary component
language. The situation in eleven-dimensional M-theory is, however,
completely different due to the fact that an off-shell lagrangian
formulation with a finite number of auxiliary fields is not known and may 
not even exist. From a general counting argument by Siegel and Ro\v cek
[\SiegelRocek] we know that this is true for $N$=4 super-Yang--Mills 
in four dimensions (and consequently also in ten dimensions)
but that maximally supersymmetric supergravity passes the test.
In higher dimensions, similar arguments were used in ref. [\TaylorRivelles]
to prove that under some mild assumptions eleven-dimensional supergravity
does in fact not allow for an off-shell lagrangian formulation. 
The analysis carried out here, when completed, will
provide an independent check of that statement.
In this respect the approach advocated here is parallel to the
discussion of ten-dimensional super-Yang--Mills theory carried out in
refs. [\BNVI] and [\BNXXIV], which does in fact prove
that an  off-shell lagrangian based on these superspace fields 
does not exist.

To implement the symmetries of any M-theory effective action in a
manifest way, we will here follow ref. [\BNLXVI] and define the theory in
superspace by means of the superspace Bianchi identities (SSBIs), which
are integrability conditions when the theory is formulated in terms of
superspace field strengths. From these
we will derive consistency conditions on the form of
the field equations. The analysis of the SSBIs will depend on the
structure of certain components of the supertorsion, and one particular 
goal is to
find connections between the various possible superinvariants and
consistent expressions for the components of the supertorsion.  The
structure of these components, as \eg\ which components can be set to
zero under which conditions, will be clarified by our analysis. This
is an important result since the torsion components are a vital input
when proving $\kappa$-invariance for M2 and M5 branes coupled to
background supergravity [\BSTMii,\DuffStelleM,\CNS] 
and M-theory corrected versions of it. In
fact, one should compare to the situation in IIA and IIB string theory
and the coupling to D-branes [\Dthree,\SchwarzDbrane,\Dp,\BTDbrane].
 Here it has been established that there
are higher-derivative background field corrections also on the world-sheets 
of the branes, see \eg\ refs. [\BBG,\Wyllard] and references therein.
The presence of such terms complicates the issue
of $\kappa$-invariance and it becomes crucial to know the exact
form of the supertorsion and to understand 
its relation to the corrections both in
target space and on the brane.

Another aspect of the higher-derivative corrections is that it is to a large
extent unclear how supersymmetry organises the infinite set of such
terms into infinite subsets unrelated by supersymmetry. From previous work
both in ten and eleven dimensions we know that adding one bosonic
$R^2$ or $R^4$ term generates an infinite set of other terms of
progressively higher order in number of derivatives. This is clear in
any on-shell theory, as discussed in detail in the type IIB case in \eg\
ref. [\GreenSethi]. In the heterotic case in ten dimensions an iterative
procedure is needed even in the superspace treatment of the theory due
to an implicit dependence on the three-form field strength in the
supercurvature that appears in the SSBI, $dH=\tr(F^2-R^2)$, used to
define the theory in superspace [\BNXXVIII \PTafs]. This situation 
resembles the one for
M-theory under discussion in this paper apart from the important fact that the
corresponding SSBI for the four-form field strength is not added as a
separate equation but will instead follow from the geometric SSBI for
the supertorsion [\CL,\HoweWeyl]. 
One might then suspect that if several linearised
superinvariants exist at any given order, some appear as a result of
iteration triggered by a lower order term, while some others, if they
appear at all, generate new infinite series of terms. Of course, 
in order to determine which series of terms do actually occur, one has to 
invoke some microscopic description of the theory or rely on a comparison 
with string theory.

In this paper we will make use of the fact that any conceivable
M-theory correction to the field equations must be compatible with
supersymmetry and local Lorentz invariance. This is built into the
SSBIs [\HoweWeyl,\BNLXVI] 
which when solved (the meaning of which is explained below)
produce constraints on the supertorsion and other superfields that
must be fulfilled by the corrections.  As a first step we prove in
this paper that the relaxed on-shell torsion constraints, argued for in
ref. [\BNLXVI], are correct and do not lead to the field equations that
follow from (\Lagrangian).  This will done without specifying the auxiliary
fields in terms of physical fields, making it possible to use the Weyl
superspace introduced by Howe [\HoweWeyl] to simplify the analysis of
the standard on-shell theory. Once the auxiliary fields are related to
physical fields, the role of Weyl superspace must be reconsidered, since
the identification will involve a dimensionful parameter ($\a'^3$ for the
$R^4$ term). This will be done elsewhere.

%\vfill\eject

\section\constraints{Relaxed torsion constraints}One of the most 
important results proved in ref. [\HoweWeyl] is that
inserting the single constraint
$$
{{T}_{\a\b}}^{c}=2{\G}_{\a\b}^{c}\komma\Eqn\onshellT
$$
used in the superspace construction of eleven-dimensional supergravity 
[\ElevenSSSG],
into the SSBIs leads to the field equations corresponding to the lowest
order lagrangian (\Lagrangian).  This constraint must therefore be relaxed in
such a way that the equations that then follow from the SSBIs are able to
accommodate any higher-derivative correction terms to the field
equations. In order to explain how this is done we need some
details of the Weyl superspace formalism. This superspace is
coordinatised by $z^{M}=(x^m,\theta^\mu)$ where $m$ enumerates the
11 bosonic and $\mu$ the 32 real fermionic coordinates
respectively. The tangent superspace has as structure group the
Lorentz group (not a superversion of it) times Weyl rescalings, and
hence one introduces a supervielbein and a superconnection 
$$
{E_{M}}^{A}(z)\komma\quad{{\O}_{MA}}^{B}(z)=
{{\o}_{MA}}^{B}(z)+{K_{MA}}^{B}(z)\komma\Eqn\superconnections
$$
where $\o_{MA}{}^B=(\o_{Ma}{}^b,{1\over4}(\G^a{}_b)_\a{}^\b\o_{Ma}{}^b)$ 
is the Lorentz part and
${K_{MA}}^{B}=(2K_{M}{\d_a}^b,{K_{M}}{\d_{\a}}^{\b})$  the Weyl
part, and the flat superindex $A=(a,\a)$ contains an SO(1,10) vector 
index $a$ and a (Majorana) spinor index $\a$. 
The super-two-form field strengths
corresponding to the fields in (\superconnections) are (suppressing the wedge
product symbol in the product of superforms) 
$$
\eqalign{
&T^{A}=DE^A=dE^A+E^B{{\O}_B}^A\komma\cr
&{R_A}^B=d{{\O}_A}^B+{\O_A}^C{\O_C}^B\komma\cr
}\Eqn\superfieldstrengths
$$
and they satisfy the SSBI
$$
\eqalign{
&DT^A=E^B{R_B}^A\komma\cr
&D{R_A}^B=0\komma\cr
}\Eqn\SSBI
$$
of which only the first one will be used in this paper. 
 Note that no separate superfield
corresponding to the four-form field strength is introduced
since both its Bianchi identity and field equation will emerge from
the analysis of the torsion SSBI. 

In order to obtain the form of the relaxed constraint given in
ref. [\BNLXVI] we expand ${{T}_{\a\b}}^{c}$ in terms of irreducible tensors
by means of the basis for symmetric $\G$-matrices
$\G^{(1)}$, $\G^{(2)}$, $\G^{(5)}$, where $\G^{(n)}$ indicate a product of
$n$ antisymmetrised $\G$-matrices with weight one, \ie,
$$
{{T}_{\a\b}}^{c}=2\left({{\G}_{\a\b}}^{d}{X_d}^c
+\fr2{{\G}_{\a\b}}^{d_1d_2}{X_{d_1d_2}}^c+
\fr{5!}{{\G}_{\a\b}}^{d_1\ldots d_5}{X_{d_1\ldots d_5}}^c\right)\komma\Eqn\aux
$$
with the understanding that the $X$'s can be further decomposed into
irreducible tensors.  We will soon discuss each one of these
irreducible $X$ tensors, but first we need to make a little digression into
what is known as ``conventional constraints'' [\ConvConstr].

Conventional constraints are used to eliminate redundant superfields
contained in the potentials of the theory, in the present case in the
vielbeins and connections. They are imposed on tangent
space tensors, here torsion components, and therefore have no 
effect on superspace reparametrisation invariance.
We can distinguish between two types of constraints. The first type makes 
use of the arbitrariness in the distinction between spin connection and 
torsion (for each irreducible representation
 contained in the spin and Weyl connections) in
eq. (\superfieldstrengths). This freedom must be fully utilised in order
to obtain a solution of the spin connection in terms of the vielbeins.
The second type of constraint makes use of arbitrary redefinitions of the 
vielbeins, $E^A\rightarrow E^BM_B{}^A$, \ie, of the choice of tangent 
bundle, while the connection is unchanged. Once again, by using the torsion
it is possible to impose conventional 
constraints in a gauge-covariant way. This second type of constraint reduces
the field content of the supervielbein, and we must clearly limit its use
to certain components in order not to lose the entire dynamics of the 
theory. An analysis  of all the possible conventional 
constraints [\BNLXVI] will leave 
only $X$'s in the 
representations 429 and 4290 in (\aux)\foot\dagger{The same statement
appears without proof in ref. [\HSW].}, as we will now discuss. 

Turning back to the components of the supertorsion at dimension 0 given in
eq. (\aux), we note that $X_d{}^c$ decomposes into representations of
dimension 65, with Dynkin label (20000), 55 with Dynkin label (01000), and
1 with Dynkin label (00000). Similarly, ${X_{d_1d_2}}^c$
goes into 11 (10000), 165 (00100) and 429 (11000), and 
${X_{d_1\ldots d_5}}^c$ into 330 (00010), 462
(00002) and 4290 (10002).
We use $M_a{}^b$ to set ${X_a}^b={\d_a}^b$ and (part of) $M_\a{}^\b$ to set all
antisymmetric tensors to zero.
At this point we
are left with the two fields which transform as 429 and 4290 under
SO(1,10). The way these appear in the supertorsion, \ie, in
${{T}_{\a\b}}^{c}$, suggests a close connection to the M2 and M5
brane, respectively, for the 429 and 4290. Although it seems easier to
deal with 429 we will in fact drop it and concentrate on the 4290
because of its probable relation to the anomaly canceling term related
to the M5 brane. (The field of interest with dimension 0, 
$\a'^3W^3+\ldots$ ($W$ is the Weyl tensor), 
does not contain the representation 429.) This will have to appear in the SSBI for the
four-form superfield strength which hence will read $d*H=\half
H^2+X_{(8)}$, where $X_{(8)}$ is the eight-form polynomial in the curvature
that was introduced in this context in ref. [\VafaWitten,\Duffoneloop]. 
 In this paper, however,
we will not take the analysis this far but instead show that the
relaxed torsion constraint [\BNLXVI]
$$
T_{\a\b}{}^{c}=2\left(\G_{\a\b}{}^{c}+
\fr{5!}\G_{\a\b}{}^{d_1\ldots d_5}X^{(4290)\,c}_{d_1\ldots d_5}\right)
\Eqn\auxbighook
$$
is general enough to lift the field equations coming from (\Lagrangian). 

More precisely, we will insert the following expanded 
irreducible torsion components
$$
\matrix{
%\hbox{dim 0:}\hfill
%&\hfill T_{\a\b}{}^{c}&=&
%2i\left(\G_{\a\b}{}^{c}
%+\fr{5!}\G_{\a\b}{}^{d_1\ldots d_5}X_{d_1\ldots d_5}{}^c\right)
%	\hfill&\hfill(10002)\cr
%\hfill&\hfill&\hfill&\hfill&\hfill\cr
\hbox{dim $\half$:}\hfill
&\hfill T_{\a b}{}^{c}&=&S_b{}^c{}_\a
	\hbox{\vbox to10pt{}}\hfill&\hfill(20001)\cr
\hfill&\hfill&\hfill&
+2(\G_{(b}S_{d)})_\a\eta^{cd}
	\hbox{\vbox to10pt{}}\hfill&\hfill(10001)\cr
\hfill&\hfill&\hfill&\hfill&\hfill\cr
\hfill&\hfill T_{\a\b}{}^{\c}&=&
\fr{120}\G^{d_1\ldots d_5}_{\a\b}Z_{d_1\ldots d_5}{}^\c
	\hbox{\vbox to10pt{}}\hfill&\hfill(00003)\cr
\hfill&\hfill&\hfill&
+\fr{24}\G^{d_1\ldots d_5}_{\a\b}(\G_{d_1}Z_{d_2\ldots d_5})^\c
	\hbox{\vbox to10pt{}}\hfill&\hfill(00011)\cr
\hfill&\hfill&\hfill&
+\fr{12}\G^{d_1\ldots d_5}_{\a\b}(\G_{d_1d_2}Z_{d_3d_4d_5})^\c
	\hbox{\vbox to10pt{}}\hfill&\hfill(00101)\cr
\hfill&\hfill&\hfill&
+\fr{12}\G^{d_1\ldots d_5}_{\a\b}(\G_{d_1d_2d_3}Z_{d_4d_5})^\c
+\fr2\G^{d_1d_2}_{\a\b}Y_{d_1d_2}{}^c
	\hbox{\vbox to10pt{}}\hfill&\hfill(01001)\cr
\hfill&\hfill&\hfill&
+\fr{24}\G^{d_1\ldots d_5}_{\a\b}(\G_{d_1\ldots d_4}Z_{d_5})^\c
+\G^{d_1d_2}_{\a\b}(\G_{d_1}Y_{d_2})^\c
	\hbox{\vbox to10pt{}}\hfill&\hfill(10001)\cr
\hfill&\hfill&\hfill&
+\fr{120}\G^{d_1\ldots d_5}_{\a\b}(\G_{d_1\ldots d_5}Z)^\c
+\fr2\G^{d_1d_2}_{\a\b}(\G_{d_1d_2}Y)^\c
	\hbox{\vbox to10pt{}}\hfill&\hfill(00001)\cr
\hfill&\hfill&\hfill&\hfill&\hfill\cr
\hbox{dim 1:}\hfill
&\hfill T_{ab}{}^{c}&=&0\hfill&\hfill\cr
\hfill&\hfill&\hfill&\hfill&\hfill\cr
}
\Eqn\TorsionIrreps
$$
$$
\matrix{
\hfill&\hfill T_{a\b}{}^{\c}&=&
\fr{24}(\G^{d_1\ldots d_4})_\b{}^\c A_{d_1\ldots d_4a}
+\fr{120}(\G_a{}^{d_1\ldots d_5})_\b{}^\c A'_{d_1\ldots d_5}
	\hbox{\vbox to10pt{}}\hfill&\hfill2(00002)\cr
\hfill&\hfill&\hfill&
+\fr6(\G^{d_1d_2d_3})_\b{}^\c A_{d_1d_2d_3a}
+\fr{24}(\G_a{}^{d_1\ldots d_4})_\b{}^\c A'_{d_1\ldots d_4}
	\hbox{\vbox to10pt{}}\hfill&\hfill2(00010)\cr
\hfill&\hfill&\hfill&
+\fr2(\G^{d_1d_2})_\b{}^\c A_{d_1d_2a}
+\fr6(\G_a{}^{d_1d_2d_3})_\b{}^\c A'_{d_1d_2d_3}
	\hbox{\vbox to10pt{}}\hfill&\hfill2(00100)\cr
\hfill&\hfill&\hfill&
+(\G^d)_\b{}^\c A_{da}
+\fr2(\G_a{}^{d_1d_2})_\b{}^\c A'_{d_1d_2}
	\hbox{\vbox to10pt{}}\hfill&\hfill2(01000)\cr
\hfill&\hfill&\hfill&
+(\G_a{}^d)_\b{}^\c A'_d
	\hbox{\vbox to10pt{}}\hfill&\hfill(10000)\cr
\hfill&\hfill&\hfill&
+(\G_a)_\b{}^\c A'
	\hbox{\vbox to10pt{}}\hfill&\hfill(00000)\cr
\hfill&\hfill&\hfill&
+\fr{120}(\G^{d_1\ldots d_5})_\b{}^\c B_{d_1\ldots d_5,a}
	\hbox{\vbox to10pt{}}\hfill&\hfill(10002)\cr
\hfill&\hfill&\hfill&
+\fr{24}(\G^{d_1\ldots d_4})_\b{}^\c B_{d_1\ldots d_4,a}
	\hbox{\vbox to10pt{}}\hfill&\hfill(10010)\cr
\hfill&\hfill&\hfill&
+\fr6(\G^{d_1d_2d_3})_\b{}^\c B_{d_1d_2d_3,a}
	\hbox{\vbox to10pt{}}\hfill&\hfill(10100)\cr
\hfill&\hfill&\hfill&
+{1\over2}(\G^{d_1d_2})_\b{}^\c B_{d_1d_2,a}
	\hbox{\vbox to10pt{}}\hfill&\hfill(11000)\cr
\hfill&\hfill&\hfill&
+(\G^d)_\b{}^\c B_{d,a}
	\hbox{\vbox to10pt{}}\hfill&\hfill(20000)\cr
\hfill&\hfill&\hfill&\hfill&\hfill\cr
\hbox{dim $3\over2$:}\hfill
&\hfill T_{ab}{}^{\c}&=&
t_{ab}{}^{\c}
	\hbox{\vbox to10pt{}}\hfill&\hfill(01001)\cr
\hfill&\hfill&\hfill&
+2(\G_{[a}t_{b]})^\c
	\hbox{\vbox to10pt{}}\hfill&\hfill(10001)\cr
\hfill&\hfill&\hfill&
+(\G_{ab}t)^\c
	\hbox{\vbox to10pt{}}\hfill&\hfill(00001)\cr
}
\eqno{\vbox{\hbox{({\old2}.{\old7})}\hbox{\it cont'd}}} 
$$
(where, in addition to the conventional constraints used at dimension 0,
we have used the ones corresponding to $\O_{MA}{}^B$ and $M_a{}^\b$
\foot\ast{In ref. [\GatesNishino], it was proposed that a spinor superfield
should be used to take the theory off-shell. This field was subsequently 
shown to be eliminated by a conventional constraint corresponding to
$\ss M_a{}^\b$ [\HoweWeyl], 
and does not occur in our expansion of the torsion.}) 
into the SSBIs and
derive the spinor part of the spin $3\over2$ field equation
$$
t_{\a}\sim(D^3X^{(4290)})_{\a}\punkt\Eqn\RSoffshell
$$
This will show that at the $\theta^3$ level in the auxiliary superfield
$X^{(4290)}$, there is a spinor field that lifts the field equation
$t_{\a}=0$ that would otherwise result from lagrangian in eq.
(\Lagrangian). See \eg\ ref. [\BNVII] for the relation between the 
Rarita-Schwinger
field equation and $t_{\a}=0$ which in 11 dimensions reads 
$\Gamma^{abc}T_{bc}=18t^a-90\Gamma^at$.

\section\solving{Solving the superspace Bianchi identities off-shell}In this 
section we give a brief account of the steps required to
derive the off-shell equation mentioned at the end of the last
section. A more complete discussion will be presented elsewhere
[\CGNNIII]. The method for solving the SSBI, $DT^A=E^B{R_B}^A$, 
is to extract its component equations and solve these by increasing
dimension.
The equation of lowest dimension, $\half$, is the one
multiplying the three-form $E^{\a}E^{\b}E^{\c}$ and with
$A=a$,
$$
0={R_{(\a\b\c)}}^d=D_{(\a}{T_{\b\c)}}^d+{T_{(\a\b}}^E{T_{|E|\c)}}^d\komma
\Eqn\dimhalfssbi
$$
where $(\ldots)$ indicates symmetrisation of the indices (except for the
ones between bars $|\ldots|$). This equation can be decomposed into a
large number of equations, each one corresponding to an irreducible tensor
appearing in the decomposition of the symmetric product of three
spinors times a vector, which is the tensor structure of the SSBI
(\dimhalfssbi). When the expansions of the torsion 
components are inserted
into these irreducible tensor equations, all irreducible tensor parts
of the torsion will drop out except the ones that
coincide with the representation specifying the equation.

We should also mention that we restrict ourselves to a linearised analysis.
A more non-linear treatment is feasible, at least in the original 
fields, but here the ordinary supergravity fields and the auxiliary ones 
are treated on equal footing. In this paper we also neglect vector
derivatives on the auxiliary superfield $X^{4290}$.

Since the equation (\dimhalfssbi) involves the fields at $\theta$ level
in $X^{(4290)}$, we need to expand also $D_{\a}{X_{a_1\ldots a_5}}^b$ into
irreducible tensors. This gives rise to the following spinorial
tensors:
$$
\matrix{
&(10001)\quad\tX_{a}^{\a}\hfill&(20001)\quad\tX_{a,b}^{\a}\hfill\cr
&(01001)\quad\tX_{a_1a_2}^{\a}\hfill&(11001)\quad\tX_{a_1a_2,b}^{\a}\hfill\cr
&(00101)\quad\tX_{a_1a_2a_3}^{\a}\qquad\hfill&\hfill\cr
&(00011)\quad\tX_{a_1\ldots a_4}^{\a}\hfill&\hfill\cr
&(00003)\quad\tX_{a_1\ldots a_5}^{\a}\hfill
	&(10003)\quad\tX_{a_1\ldots a_5,b}^{\a}\hfill\cr
}
\Eqn\xone
$$
where the $\tX$'s with $a_1\ldots a_n$ are $\G$-traceless $n$-forms and
the ones with $a_1\ldots a_n,b$ are $\G$-traceless tensors with $n$ 
antisymmetric indices
and $\tX_{[a_1\ldots a_n,b]}=0=\tX_{a_1\ldots a_n,b}\eta^{a_nb}$ 
(note the distinction
between these two kinds of representation, kept track of by the
comma between the $a_i$ indices and the $b$ index).  The solution to
the SSBI of dimension $\half$ then relates the various irreducible
tensors in the supertorsion to the $\tX$'s in
(\xone). 
In fact, as a consequence of using Weyl superspace, with the
extra conventional constraints associated with the Weyl connection, we find
that the torsions involved in this SSBI are uniquely
determined by the $\tX$'s above. Thus if we set $X^{(4290)}$ to zero these
torsions will vanish without invoking any extra assumptions, a result
that also follows from the work of Howe in ref. [\HoweWeyl]. Besides these
relations, the solution also tells us that $\tX_{a,b}^{\a}$ and
$\tX_{a_1\ldots a_5,b}^{\a}$ are zero (the former one only when the 
representation 429 is left out of the dimension 0 torsion).
In this paper, we spare the reader from the exact expression for the 
torsion components in terms of $X^{(4290)}$. The calculation involves a 
certain degree of technical complexity, which is left for ref. [\CGNNIII]. 

We now turn to the SSBIs with dimension 1. There are two such
equations, namely 
$$
\eqalign{
&R_{\a\b c}{}^d=2D_{(\a}T_{\b)c}{}^d+D_cT_{\a\b}{}^d
	+T_{\a\b}{}^ET_{Ec}{}^d+2T_{c(\a}{}^ET_{|E|\b)}{}^d\komma\cr
&R_{(\a\b\c)}{}^\d=D_{(\a}T_{\b\c)}{}^\d+T_{(\a\b}{}^ET_{|E|\c)}{}^\d\punkt\cr
}
\Eqn\DimOneBI
$$
To deal with these equations, we must expand the superfield $X^{(4290)}$
at the $\theta^2$ level, and take into account the results already
obtained at $\theta$ level. As mentioned above, we aim at showing that the
introduction of the auxiliary fields generates a right-hand side of the
spinor part of the equation of motion for the gravitino field. 
At present, we therefore only need to consider the irreducible tensors
at dimension 1 whose spinorial derivative contains a spinor.
These are the forms, denoted $A$ and $A'$ in eq. (\TorsionIrreps).
Since now the curvatures entering eq. (\DimOneBI) are non-zero taking values in
the structure group, they must be eliminated. From the $R_{\a\b c}{}^d$
we get the information that the symmetric traceless part in $cd$ has to vanish,
and the rest are used to eliminate $R_{\a\b\c}{}^\d$ by the structure
group condition. In contrast to the equations at dimension $\half$, where
the full representation content of the index structure of the SSBI made
impact on the fields (to the extent that the representations were present
at level $\theta$ in $X^{(4290)}$), some equations now turn out to be
linearly dependent. A naive counting of fields and equations fails, and,
as we will see,
this is absolutely essential in order for the auxiliary superfield to contain
components entering the equations of motion.
This exceptional behaviour relies on the exact form of the 
solutions at dimension $\half$,
and comes at work for the three-forms, where three
equations reduce to two, and for the four-forms, where all three equations
are identical. 
The zero-, one-, two- and five-forms at second level in $X$
are set to zero (modulo terms $\sim D_bX_{a_1\ldots a_5}{}^b$ in
the five-forms), and the relevant surviving part is parametrised as
$$
\eqalign{
\fr{10}&D_{[\a}D_{\b]}X_{a_1\ldots a_5,b}				\cr
&=\G_{[a_1a_2a_3}{}^eV_{a_4a_5]be}
	+\G_{b[a_1a_2}{}^eV_{a_3a_4a_5]e}
	-\Fr67\eta_{b[a_1}\G_{a_2a_3}{}^{e_1e_2}V_{a_4a_5]e_1e_2}	\cr
&+\G_{a_1\ldots a_5}{}^{e_1e_2e_3}W_{be_1e_2e_3}
	+\G_{b[a_1\ldots a_4}{}^{e_1e_2e_3}W_{a_5]e_1e_2e_3}
	-\Fr67\eta_{b[a_1}
		\G_{a_2\ldots a_5]}{}^{e_1\ldots e_4}W_{e_1\ldots e_4}	\cr
&+\G_{[a_1a_2a_3}V_{a_4a_5]b}
	+\G_{b[a_1a_2}V_{a_3a_4a_5]}
	-\Fr67\eta_{b[a_1}\G_{a_2a_3}{}^eV_{a_4a_5]e}			\cr
&+\G_{a_1\ldots a_5}{}^{e_1e_2}W_{be_1e_2}
	+\G_{b[a_1\ldots a_4}{}^{e_1e_2}W_{a_5]e_1e_2}
	-\Fr67\eta_{b[a_1}\G_{a_2\ldots a_5]}{}^{e_1e_2e_3}W_{e_1e_2e_3}\cr
&+\ldots\cr
}
\Eqn\DtwoX
$$
with the relations
$$
A+2A'+{2\ggr5\over7\ggr11\ggr23}\left(547\,V+2^5\ggr3^3\ggr17\,W\right)=0
\Eqn\DimOneFourFormEq
$$
for the four-forms, and
$$
\eqalign{
&A=-{2^2\over3\ggr11\ggr23}\left(89\,V+2^2\ggr3\ggr5\ggr139\,W\right)\komma\cr
&A'={2^3\over3\ggr11\ggr23}\left(2\ggr47\,V-3\ggr5\ggr41\,W\right)\cr
}\Eqn\DimOneThreeFormEq
$$
for the three-forms. The linear dependence already mentioned makes us
confident in these expressions. 
In the unmodified supergravity ($V=W=0$), one four-form 
in the dimension 1 torsion
survives, and is identified with the four-form field strength $H$.
In the present situation, it is a priori not obvious which combination
of the three surviving four-forms that should be identified with this
physical field (the criterion being that it is closed), and the answer to
this question will have to await the solution of the SSBIs at dimension 2.

One further result at dimension 1 is that the Weyl part of the curvature,
$G_{\a\b}=\fr{32}R_{\a\b\c}{}^\c$, vanishes, 
as was shown in ref. [\HoweWeyl] for
the unmodified supergravity. 
This is a very positive sign, since it
indicates that the theory even with the relaxed torsion constraints 
we have used to
take it off-shell is equivalent to one
with only the Lorentz group as structure group as discussed by Howe 
[\HoweWeyl]. We will comment on this further in 
the concluding section.

Finally, we consider the SSBIs at dimension $\Fr32$, which read
$$
\eqalign{
&2R_{\a[bc]}{}^d=D_\a T_{bc}{}^d+2D_{[b}T_{c]\a}{}^d
	+2T_{\a[b}{}^ET_{|E|c]}{}^d+T_{bc}{}^ET_{E\a}{}^d\cr
&2R_{a(\b\c)}{}^\d=D_a T_{\b\c}{}^\d+2D_{(\b}T_{\c)a}{}^\d
	+2T_{a(\b}{}^ET_{|E|\c)}{}^\d+T_{\b\c}{}^ET_{Ea}{}^\d\cr
}
\Eqn\DimThreeHalvesBI
$$
In an unconstrained superfield in the representation 4290, there are
two spinors at level $\theta^3$. The index structure of the SSBIs at this
level also contains two spinor equations. We have to take into account what
has been learned about $X^{(4290)}$ at lower levels. Specifically, 
at dimension 1 some of the antisymmetric tensors containing a spinor at
the next level vanished. Miraculously, again, all of these go into the same
linear combination, while the spinor coming from the three- and four-forms
survives. One spinor thus remains, and goes into part of the field equation
for the Rarita--Schwinger field, which then reads
$$
t_\a={17\over2^2\ggr3^3 \ggr 5^2 \ggr 7 \ggr 11 \ggr 13 \ggr 61}
(\G^{b_1b_2b_3}){}^{\b\c}(\G^{b_4b_5a}){}_\a{}^\d
	D_{[\b}D_\c D_{\d]}X_{b_1\ldots b_5,a}
\Eqn\NewRSEq
$$
Also the spinor component of the Weyl curvature, 
$G_\a=\fr{32}(\G^a)_\a{}^\b R_{a\b\c}{}^\c$, is set to zero.

These calculations involve some rather heavy $\Gamma$-matrix algebra which
has been facilitated enormously by the development of a Mathematica based 
program [\Ulf]. In particular, the results in this paper rely on a large
number of Fierz identities which, as explained below, can be 
completely systematised. By using the algebraic program to compute some
small number of final coefficients, any computation requiring 
Fierzing is easily dealt with. The systematisation and use of the program
is most clearly explained through an example: Consider the two spinors
appearing at $\theta^3$ level in the superfield $X^{4290}$. 
We would like to know how these two spinors are related 
via supersymmetry to the
three- and four-forms at level $\theta^2$. For this one needs to discuss
the Fierz identities for structures of the kind 
$(\Gamma^{ab...})_{[\a\b}(\Gamma^{cd...})_{\gamma]\d}$ where the vector indices
define the representation 4290 and $[...]$ means antisymmetrisation.
There are exactly seven such structures differing in the way the vector 
indices are distributed. However, since the representation 4290 occur exactly
twice in the product of a spinor with three antisymmetrised spinors, there must
be precisely five {\it linearly independent} Fierz identities 
between the seven $\G\G$ structures. Or in other words, by picking two of the
structures as independent, the others are expressible as linear combinations 
of these. The corresponding coefficients are then computed by 
contracting these relations
with a basis of $\Gamma$ matrices using the algebraic
program to perform the resulting algebra. Using this technique it is  
easy to disentangle
all the information hidden in the various SSBIs discussed in this paper.

\section\conclusions{Conclusions}We have demonstrated that, through a series
of seemingly miraculous numerical coincidences (which, however, due to the 
similarities with ten dimensions [\BNXXXI], both regarding the constraints and the subsequent manipulations of the SSBIs, one would strongly expect to occur),
 the relaxation of the torsion constraint 
at dimension zero is capable of accommodating an off-shell formulation.
By off-shell we here simply mean that the equations of motion from 
(\Lagrangian) are  
relaxed by the introduction of a current supermultiplet, contained in the
supertorsion along with the supergravity multiplet.
In this sense, the term ``on-any-shell'' might be more appropriate. 
However, since the non-existence of an off-shell action has to our
knowledge not been proven, the possibility is not ruled out that the auxiliary
fields produced by this formalism are the correct ones for the construction
of such an action.
We would like to stress that since the degrees of freedom contained in
the eleven-dimensional supergravity multiplet describe only low-energy
effective dynamics of M-theory, and this system is not supposed to be subject
to quantisation, the absence of an action at this level is completely
acceptable.
The results are sofar partial. We have not yet investigated all equations
of motion. In a following paper [\CGNNIII], 
we will give a more detailed account
of the calculations.

An obvious application of the formalism, as mentioned in the introduction, 
is to use it to derive higher-derivative corrections to M-theory, beginning
with $R^4$ terms and their superpartners. 
The identification of our auxiliary field $X^{(4290)}$ as a supergravity
self-interaction clearly breaks Weyl invariance. It is encouraging to note
that the corresponding curvatures vanish, as far as our analysis goes,
which indicates that the correct procedure is to restrict to the Lorentz
structure group in order to avoid ambiguities in the definition of
Weyl weights, while retaining the corresponding conventional constraints.

Brane dynamics in general backgrounds is most conveniently described in 
terms of quantities pulled back from target superspace to the world-volume.
It is known that $\kappa$-symmetry quite generally demands the background
fields to be on-shell. This must still be true for branes in backgrounds
modified by higher-derivative corrections. We believe that our formalism 
will be essential for such an analysis. One question arises directly: Is 
the action for \eg\ the M2-brane still given by the same expression,
$$
S\sim\int d^3\sigma\sqrt{-g}+\int C\komma\Eqn\MTwoAction
$$
so that the corrections come only through the pullbacks of the modified
background fields, or is this form changed?
We have not discussed the superspace tensor fields in this paper, but by 
analysing the dimension zero identity, one
realises that the equation $dH=0$ demands $H$ to have non-vanishing
components even at negative dimensions. These will appear in a 
$\kappa$-transformation of the WZ term in eq. (\MTwoAction), but do not have
any torsion counterpart to cancel. 
We hope to be able to come back also to this issue. 

Finally, it 
would also be interesting to investigate in a strict sense whether the
assumption of locality, which is implicit in our work, limits the current
multiplet to self-interactions of the supergravity multiplet or whether
there are traces of interactions with other M-theory states.

\acknowledgements
The authors are grateful to Kasper Peeters, Pierre Vanhove and Anders
Westerberg for very inspiring discussions and for their generous
communication of work in progress [\PVW]. This work is partly  
supported by EU contract HPRN-CT-2000-00122 and by the Swedish and Danish 
Natural Science
Research Councils.

\noindent For some representation theoretical considerations, the program 
LiE [\LiE] has been useful.

%\vfill\eject
\xrm\refout 

\end